\documentclass[twocolumn,english,prl,amsmath,amssymb]{revtex4}
\usepackage[T1]{fontenc}
\usepackage[latin1]{inputenc}
\usepackage{graphicx}
\usepackage{epstopdf}
\usepackage{amssymb}
\usepackage{amsmath}
\usepackage{dcolumn}
\usepackage{color}
\usepackage{bm}
\usepackage{babel}
\usepackage[normalem]{ulem}
\usepackage{booktabs}
\usepackage{multirow}
\usepackage{indentfirst}
\usepackage{pstricks}
\usepackage{verbatim}

\begin{document}

\title{Ultrafast and widely tuneable vertical-external-cavity surface-emitting laser, mode-locked by a graphene-integrated distributed Bragg reflector}
\author{C.A. Zaugg$^1$, Z. Sun$^{2}$, V.J. Wittwer$^{2}$, D. Popa$^2$, S. Milana$^2$,\\ T. S. Kulmala$^2$, R.S. Sundaram$^2$, M. Mangold$^1$, O.D. Sieber$^1$,\\ M. Golling$^1$, Y. Lee$^3$, J.H. Ahn$^3$, A.C. Ferrari$^2$, U. Keller$^1$}

\affiliation{$^1$ Department of Physics, Institute for Quantum Electronics, ETH Z\"{u}rich, Wolfgang-Pauli-Str. 16, 8093 Z\"{u}rich, Switzerland\\
$^2$ Cambridge Graphene Centre, University of Cambridge, Cambridge CB3 0FA, UK\\
$^3$ School of Electrical \& Electronic Engineering, Yonsei University, Seoul 120-749, Korea}

\begin{abstract}
We report a versatile and cost-effective way of controlling the unsaturated loss, modulation depth and saturation fluence of graphene-based saturable absorbers (GSAs), by changing the thickness of a spacer between SLG and a high-reflection mirror. This allows us to modulate the electric field intensity enhancement at the GSA from 0 up to 400\%, due to the interference of incident and reflected light at the mirror. The unsaturated loss of the SLG-mirror-assembly can be reduced to$\sim$0. We use this to mode-lock a VECSEL from 935 to 981nm. This approach can be applied to integrate SLG into various optical components, such as output coupler mirrors, dispersive mirrors, dielectric coatings on gain materials. Conversely, it can also be used to increase absorption (up to 10\%) in various graphene based photonics and optoelectronics devices, such as photodetectors.
\end{abstract}
\maketitle

Ultrafast mode-locked lasers play an increasingly important role in numerous applications, ranging from optical communications\cite{Keller_nature_03} to medical diagnostics\cite{Keller_pr_06} and industrial material processing\cite{Martin_book}. In particular, ultrafast vertical-external-cavity surface-emitting lasers (VECSELs), also referred to as semiconductor disk lasers (SDLs)\cite{Okhotnikov_book} or optically pumped semiconductor lasers (OPSLs)\cite{Keller_nature_03,Keller_pr_06,Okhotnikov_book}, are excellent pulsed sources for various applications, such as multi-photon microscopy\cite{Aviles_boe_11}, optical data communications\cite{Okhotnikov_book}, supercontinuum generation\cite{Wilcox_oe_13} and ultra-compact stabilized frequency combs \cite{Keller_pr_06,Okhotnikov_book}. In such lasers, light propagates perpendicular to the semiconductor gain layers\cite{Okhotnikov_book}. In contrast to vertical-cavity surface-emitting lasers (VCSELs)\cite{Jewell_JQE_91}, a VECSEL consists of an external cavity, formed by high-reflection mirrors, and an output coupler, with typical cavity lengths of a few mm up to tens cm\cite{Keller_pr_06,Keller_nature_03}. The gain chip generally contains a highly reflective bottom section to reflect the laser and pump light, an active semiconductor gain section in the middle, and an anti-reflective top layer\cite{Keller_pr_06,Keller_nature_03,Okhotnikov_book}. VECSELs combine the advantages of semiconductor laser technology, such as compact footprint (down to$\sim$3mm cavity length\cite{Lorenser_JQE_06}), with the benefits of diode pumped solid-state lasers, such as low timing jitter\cite{Wittwer_p_11}, excellent beam quality\cite{Rudin_oe_10}, high average\cite{Rudin_oe_10,Heinen_el_12} and peak power\cite{Wilcox_oe_13,Scheller_el_12}.

Currently, semiconductor saturable absorber mirrors (SESAMs)\cite{Keller_nature_03} are used for passive mode-locking, since they offer advantages such as an excellent ratio of saturable to non-saturable losses (e.g.50:1\cite{Saraceno_jstqe_12}) and a high damage threshold ($>$0.21J/cm$^2$)\cite{Saraceno_jstqe_12}. However, SESAMs, epitaxially grown on lattice-matched semiconductor substrates\cite{Keller_nature_03}, only offer a limited operation bandwidth (to date, the broadest tuning range of VECSELs mode-locked with SESAMs is 13.7nm\cite{Morris_pw_12}) and have a fast recovery time ranging from several hundreds fs\cite{Hoffmann_oe_11} to tens ps\cite{Saraceno_jstqe_12}. Graphene, on the other hand, is the widest bandwidth material\cite{Bonaccorso_np_10}, due to the gapless linear dispersion of the Dirac electrons, and has ultrafast recovery dynamics ($<$100fs)\cite{Brida_nc_13,Tomadin_prb_13}. Furthermore, large-area (compared to a typical laser spot), high quality, single layer graphene (SLG) can be easily grown\cite{Bonaccorso_mt_12} and integrated in a variety of lasers\cite{Bonaccorso_np_10,Sun_pe_12}. Graphene has emerged as a promising saturable absorber (SA) for ultrafast pulse generation because of its simple, low-cost fabrication and assembly\cite{Bonaccorso_np_10,Hasan_am_2009,Sun_an_10}, ultrafast carrier lifetime\cite{Brida_nc_13,Tomadin_prb_13} and broadband absorption\cite{Bonaccorso_np_10,Nair_s_08,Mak_prl_08}. The unsaturated loss (i.e. the loss of a device at low incident power) of a typical intracavity transmission device based on single layer graphene (SLG) is typically $\sim$2$\times$2.3\% (the factor 2 accounting for the double-pass per round-trip) for the most common linear cavities\cite{Baek_ape_2012,Lag_apl_13}. While this allows to use SLG as SA (GSA) to mode-lock a variety of lasers, such as fiber\cite{Hasan_am_2009,Sun_an_10}, solid-state\cite{Bonaccorso_np_10,Lag_apl_13} and waveguide\cite{Rose_oe_13}, it poses serious limitations for VECSELs\cite{Keller_pr_06}. These typically require a SA mirror with losses$<$3\%\cite{Mangold_oe_12} because the small-signal gain (i.e. the optical gain for a low-intensity signal where no saturation occurs during amplification) of VECSELs suitable for mode-locking is$\sim$3 to 5\%\cite{Mangold_oe_12}. Thus, inserting a SLG-based device (e.g. SLG on a quartz substrate\cite{Lag_apl_13}) inhibits lasing, due to the high loss induced by the$\sim$4.6\% absorption incurred in the double-pass per cavity round-trip.
\begin{figure*}[ht!]
\centerline{\includegraphics[width=180mm]{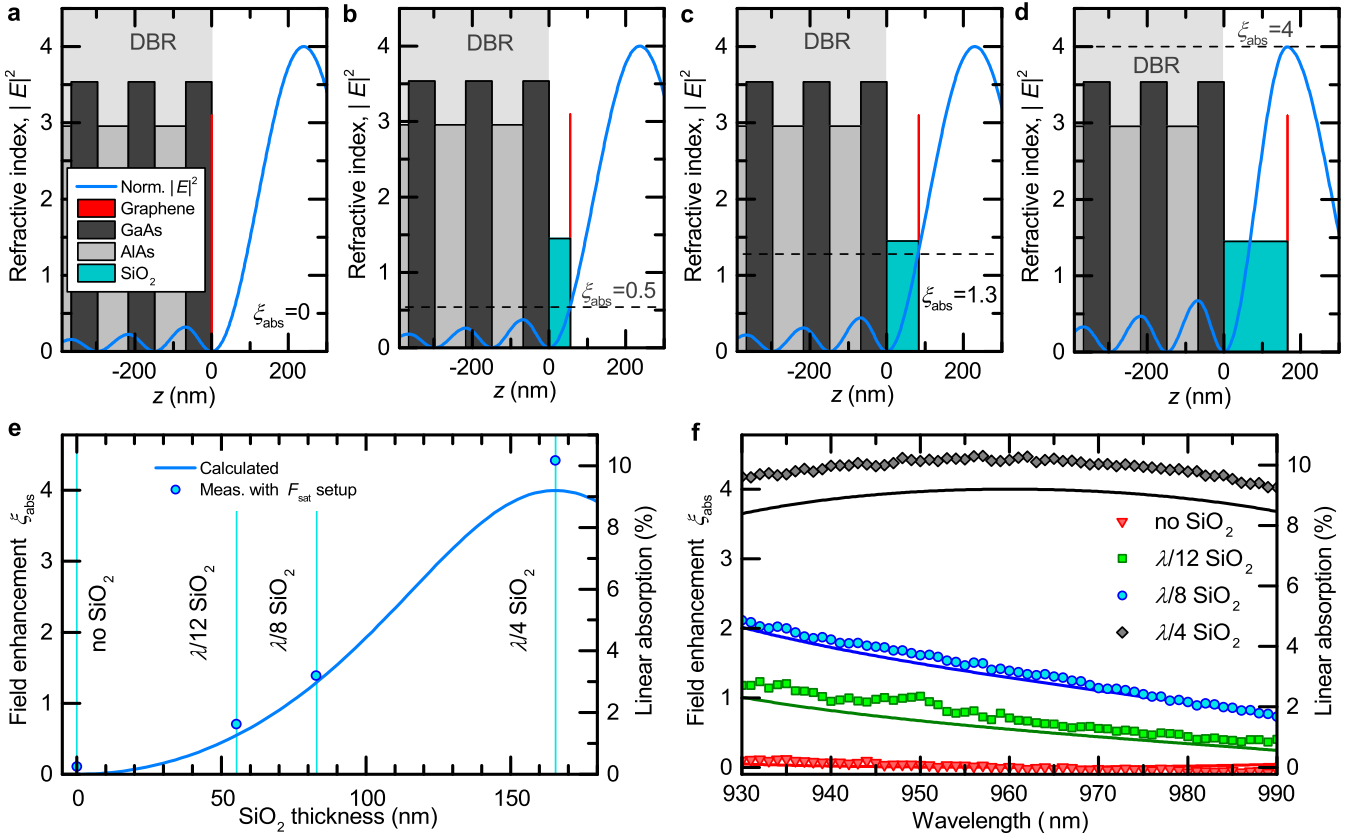}}
\caption{\label{figure1} DBR-GSAM design. Schematic zoom into the last mirror pairs with (a) no SiO$_2$, (b) $\lambda$/12 (55 nm) SiO$_2$, (c) $\lambda$/8 (83 nm) SiO$_2$ and (d) $\lambda$/4 (165 nm) SiO$_2$. The blue curve represents the normalized standing electric field intensity resulting from the refractive index profile, as a function of the vertical distance from the mirror, for the design wavelength $\lambda$=960nm. A SLG is placed as the last layer. (e) (left axis) linear absorption and (right axis) field intensity enhancement at the SLG location corresponding to the DBRs without SiO$_2$ ($\xi_{\text{abs}}$=0), a $\lambda$/12 layer of SiO$_2$ ($\xi_{\text{abs}}$=0.5), a $\lambda$/8 layer ($\xi_{\text{abs}}$=1.3) and a $\lambda$/4 layer ($\xi_{\text{abs}}$=4). (f) (lines) calculated and (dots) experimental $\xi_{\text{abs}}$ and absorption of the four designs as a function of wavelength.}
\end{figure*}
\begin{figure*}
\centerline{\includegraphics[width=180mm]{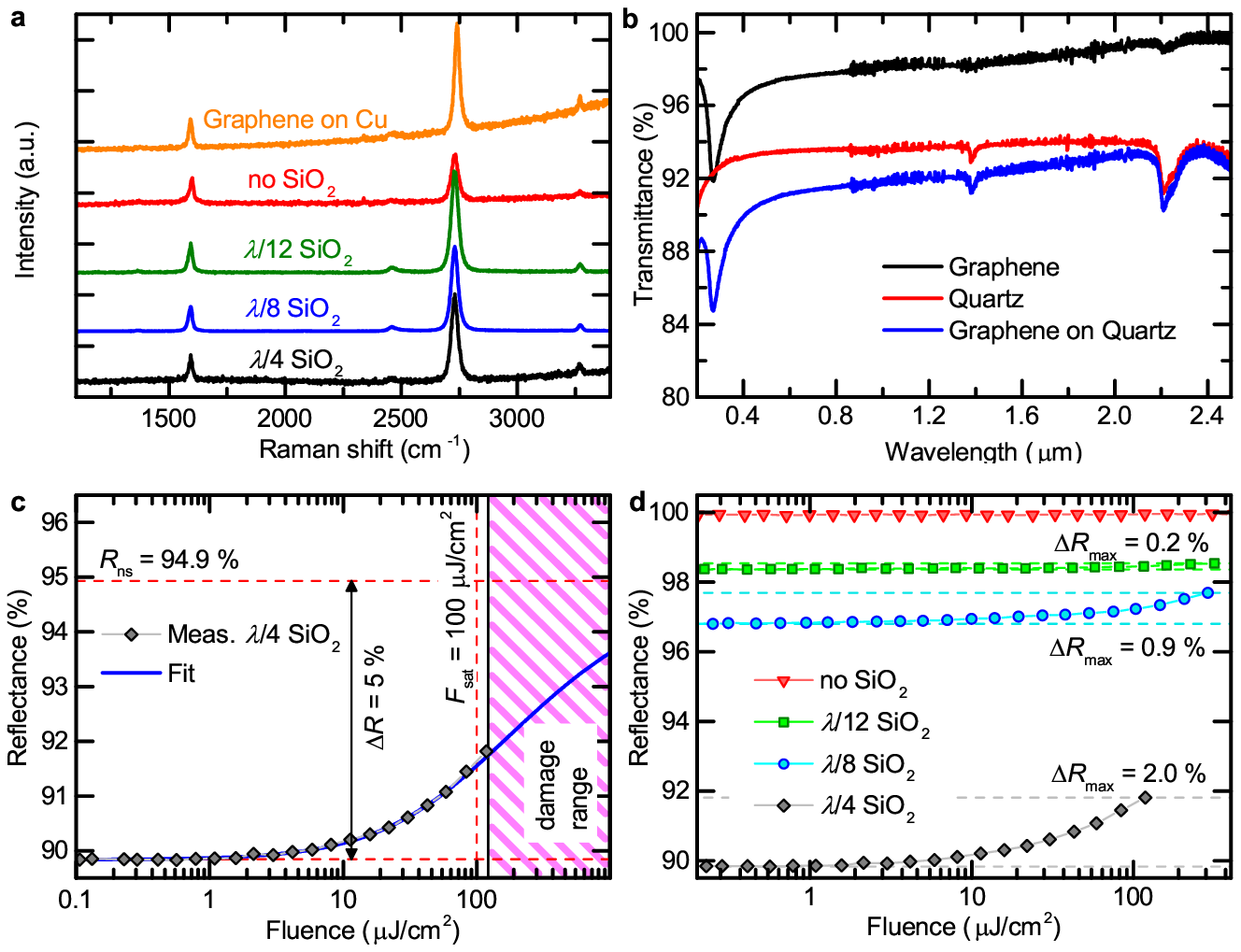}}
\caption{\label{figure2} Raman characterization and non-linear response.(a) Raman spectra of graphene on copper and after transfer on the no-SiO$_2$, $\lambda$/4 SiO$_2$, $\lambda$/8 SiO$_2$, $\lambda$/12 SiO$_2$ devices. (b) Transmittance of the same graphene transferred on a quartz substrate, derived from the transmittance of transferred graphene on quartz divided by that of quartz. (c) Non-linear reflectivity of the $\lambda$/4 SiO$_2$ sample (black markers) and fit assuming a 5\% saturable and 5.1\% non-saturable absorption (blue curve), resulting in a saturation fluence of 100 $\mu$J/cm$^2$. (d) Non-linear reflectivity of all GSAMs.}
\end{figure*}

To realize VECSEL mode-locking with graphene it is thus crucial to reduce the losses per cavity roundtrip to$<$3\% (i.e.$<$1.5\% for single pass) while maintaining high (in the range of 0.5-2\%\cite{Okhotnikov_book}) modulation depth (i.e. the maximum absorption change induced by changing the intensity of the incident light) over a spectral range wide enough to have a sufficient modulation for the self-starting passive mode-locking of broadband VECSELs. Different methods can be used to reduce the absorption in graphene: Doping\cite{Mak_prl_08,Lee_apb_2012} or gating\cite{Wang_s_08} can decrease the absorption over a broad spectral range by Pauli blocking according to\cite{Mak_prl_08,Lag_apl_13}: $A\left({\lambda,T}\right)=\frac{\pi^2 e^2}{hc}\left[{\text{tanh}\left(\frac{\frac{hc}{\lambda}+2E_{\text{F}}}{4k_\text{B}T}\right)+\text{tanh}\left(\frac{\frac{hc}{\lambda}-2E_{\text{F}}}{4k_\text{B}T}\right)}\right]$, where T is the temperature and $E_{\text{F}}$ is the Fermi level. So, e.g., to have 1.5\% absorption at$\sim$960nm (the working wavelength of our laser) one would need to stably shift the Fermi level by$\sim$630meV. However, it is challenging to precisely control this high doping level. Gating usually needs extra electrical contacts and drivers, which increase the complexity of the system.

Here, we change the absorption by controlling the electric field intensity in SLG on a high-reflection mirror. The resulting SLG-based saturable absorber mirrors (GSAMs) have an unsaturated loss adjustable from 0 up to 10\% and modulation depth up to 5\%. These enable us to mode-lock a VECSEL, at the same time exploiting the broadband properties of graphene, thus allowing the widest wavelength-tuning thus far reported in VECSELs.
\begin{figure*}[ht!]
\centerline{\includegraphics[width=180mm]{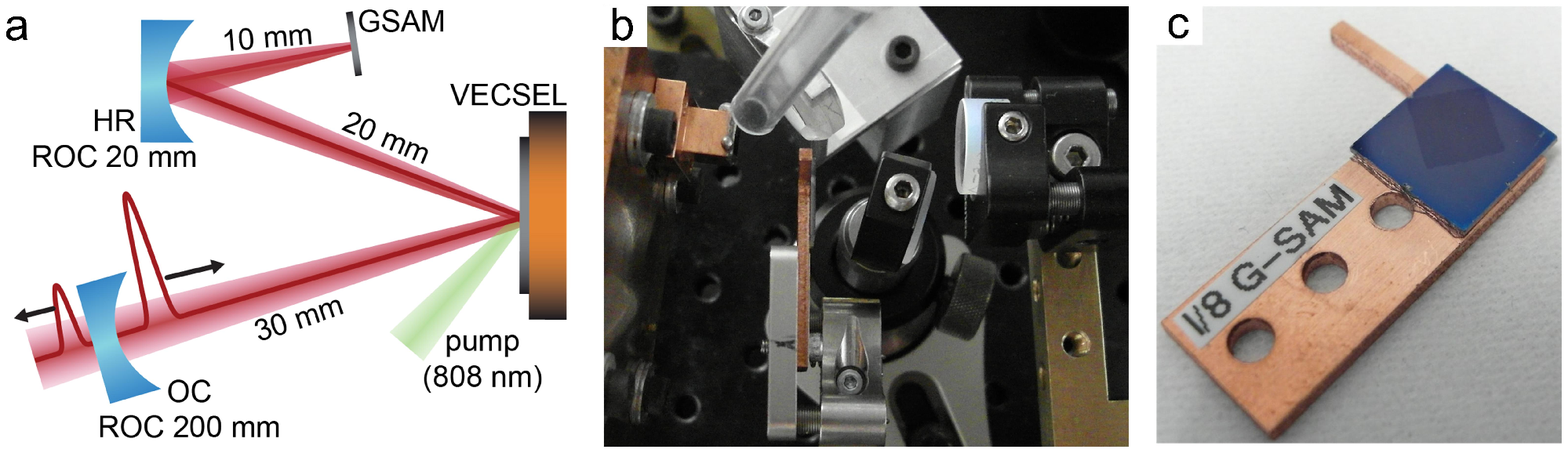}}
\caption{\label{figure3} Laser setup. (a) Schematic of the VECSEL setup. OC: output coupler mirror. HR: high reflective folding mirror. GSAM: graphene saturable absorber. The VECSEL gain chip is placed as a folding mirror and pumped under a 45$^\text{o}$ angle. The total cavity length is 6cm. (b) Picture of the laser setup. (c) Picture of the $\lambda$/8 GSAM. The SLG is clearly seen as shaded area, since the 83nm SiO$_2$ thickness gives a a high optical contrast in the visible range\cite{Cas_nl_07}.}
\end{figure*}

The GSAM absorption is controlled as follows. The incoming and reflected waves off a mirror, form a standing wave beyond the mirror surface. The field intensity enhancement $\xi (z)$ at a distance $z$ from the mirror can be written as\cite{Lee_apb_2012,Spuhler_apb_05}:
\begin{equation}\label{equationFEdef}
\xi\left({z}\right)=\frac{ \left| \mathcal{E}_{\text{in}}\left({z}\right) +\mathcal{E}_{\text{out}}\left({z}\right) \right|^2}{\left|\mathcal{E}_{\text{in}}\left({z}\right)\right|^2},
\end{equation}
where $\mathcal{E}_{\text{out}}$ and $\mathcal{E}_{\text{in}}$ are the reflected and incident wave electric fields. For an anti-resonant high-reflection ($\sim100\%$) mirror with no additional coating, we get (see Methods):
\begin{equation}\label{eq_standingwave1}
\xi (z)\approx 4 \sin\left(\frac{2 \pi n_{\text{air}} z}{\lambda}\right)^2,
\end{equation}
where $\lambda$ is the wavelength, $n_{\text{air}}$ is the refractive index of air. Therefore, the SLG absorption can be tuned by changing the optical distance between SLG and the mirror surface. The SLG absorption (A) becomes $A=\alpha \xi_{\text{abs}}$, where $\alpha\sim2.3\%$ is the absorption of a suspended and undoped SLG\cite{Nair_s_08}, and $\xi_{\text{abs}}$ is the field intensity enhancement at the absorber position. E.g., placing a SLG directly onto the mirror surface ($z=0$nm) we get $\xi_{\text{abs}}=0$, thus expect no absorption, due to destructive interference between incoming and reflected waves. If SLG is placed at a $\lambda /4$ distance, where there is a peak of the standing wave, we have $z=\lambda /4$, $\xi_{\text{abs}}=4$. Thus its absorption will increase to 400\% (i.e. 4$\times 2.3\%\sim9.2\%$) due to constructive interference.
\begin{figure*}
\centerline{\includegraphics[width=180mm]{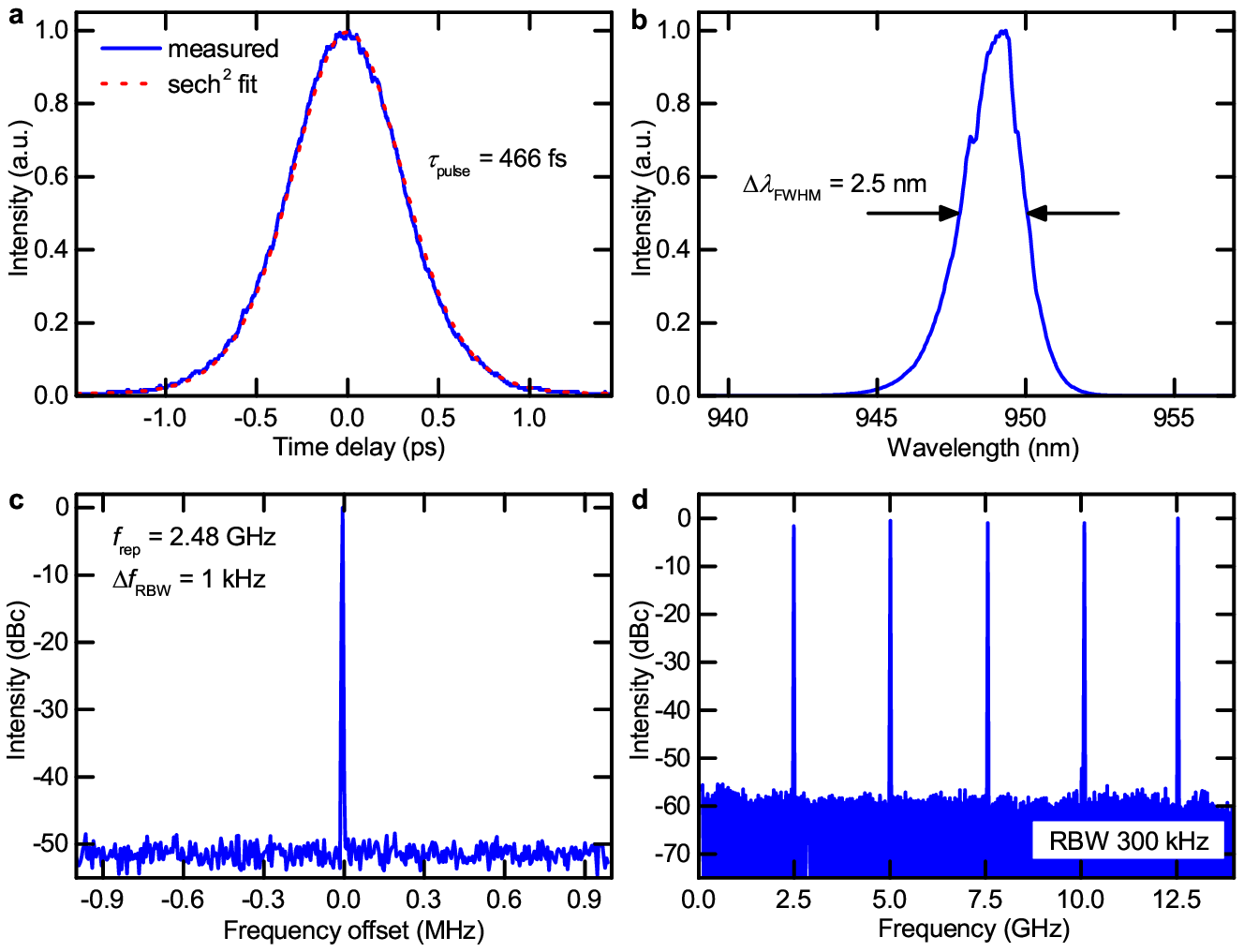}}
\caption{\label{figure4} Mode-locking results. (a) (Blue line) second harmonic autocorrelation signal and (dashed red line) fit with the autocorrelation of an ideal sech$^2$-shaped pulse, corresponding to a pulse duration of 466 fs. (b) Optical spectrum. (c) Microwave spectrum centered around the repetition rate of 2.5 GHz, measured with a 1kHz resolution bandwidth (RBW). (d) Microwave spectrum measured from 0 to 13 GHz with RBW=300kHz, showing the first 5 harmonic peaks of the repetition rate ($f_\text{rep}$).}
\end{figure*}
\begin{figure*}
\centerline{\includegraphics[width=180mm]{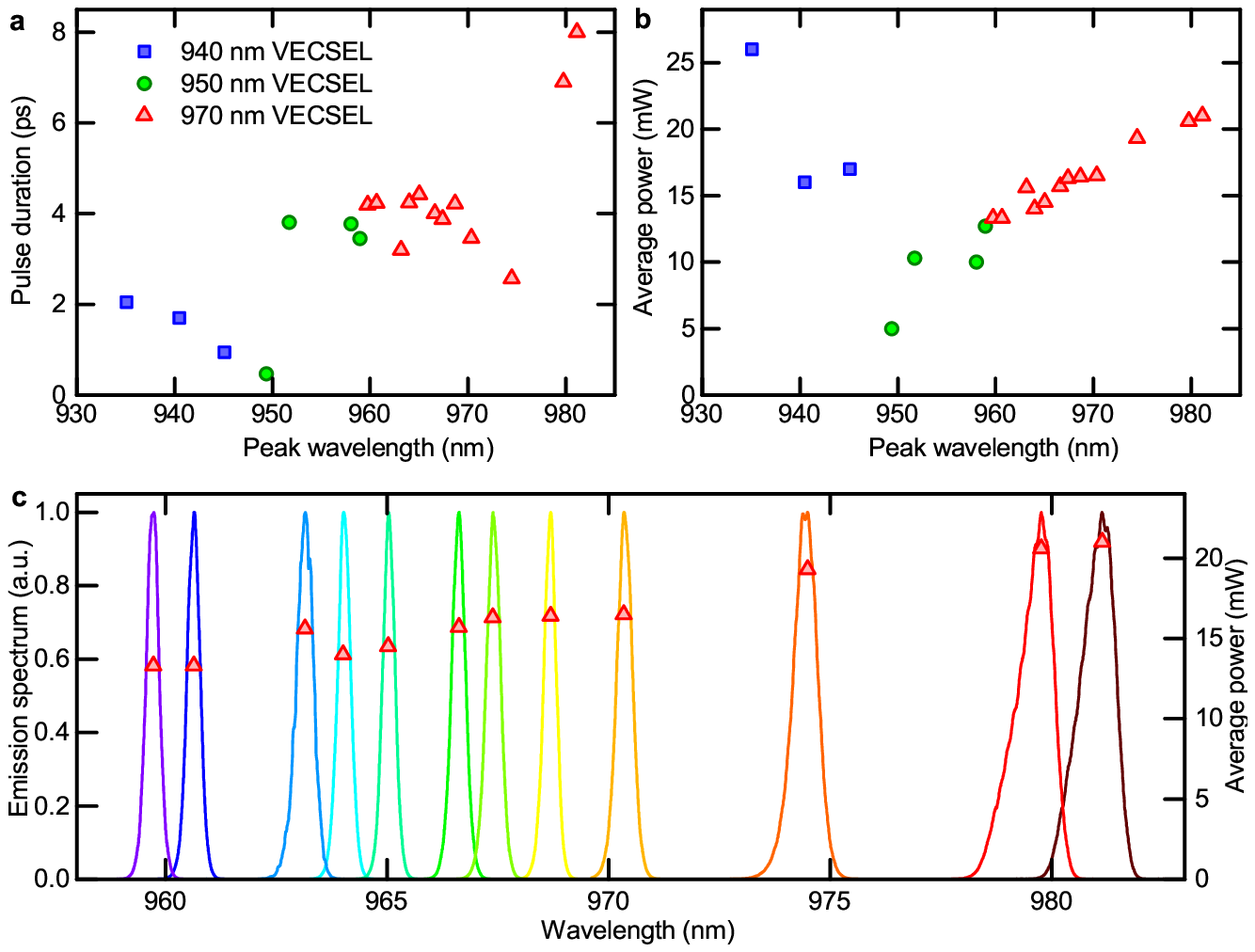}}
\caption{\label{figure5} Mode-locking with the $\lambda$/8 GSAM in VECSELs optimized for different emission wavelengths. An intra-cavity etalon is used, except for the two points at 935 and 949nm. (a) Pulse duration and (b) average output power at different emission wavelengths. (c) Emission spectra for the 970nm-VECSEL and average output power.}
\end{figure*}

To verify this principle, we fabricate four GSAMs with different optical distances by coating the mirror with: 0, $\lambda$/12 SiO$_2$, $\lambda$/8 SiO$_2$ and $\lambda$/4 SiO$_2$. We use anti-resonant distributed Bragg reflectors (DBRs)\cite{Renk_book,Spuhler_apb_05} as high-reflection mirrors. These typically consist of a stack of multiple layers with alternating high and low refractive index\cite{Renk_book,Spuhler_apb_05}, each with an optical thickness of a quarter of the design wavelength. The partial reflections at the layer interfaces can interfere constructively resulting in high reflection ($\sim100\%$\cite{Renk_book,Spuhler_apb_05}). Our DBRs consist of 30 layer pairs of AlAs and GaAs (as described in Methods). They are designed to give a node of the standing wave at the surface of the top layer (anti-resonance), with reflectivity>99.997\% at 960nm (our VECSELs wavelength). Figs.\ref{figure1}(a-d) plot schematics of the DBR. The DBRs are then coated by plasma enhanced chemical vapor deposited SiO$_2$ with different thickness ($d_{\text{SiO}_{2}}$): 0, $\lambda$/12, $\lambda$/8 and $\lambda$/4, i.e 0, 55, 83, 165nm. The field intensity enhancement $\xi$ at the top surfaces of these devices can be calculated as (see Methods):
\begin{equation}\label{equationFE_abs}
\xi_{\text{abs}} \approx \frac{4}{1+ n_{\text{SiO}_{2}}^2\cot^2\left(\frac{2\pi}{\lambda} n_{\text{SiO}_{2}} d_{\text{SiO}_{2}}\right)},
\end{equation}
where $ n_{\text{SiO}_{2}}$ and $d_{\text{SiO}_{2}}$ are the refractive index and the thickness of the SiO$_2$ spacers at the operation wavelength $\lambda$. This gives a field intensity enhancement $\xi_{\text{abs}}$ of 0, 0.5, 1.3 and 4 respectively.

SLG is then grown by chemical vapor deposition (CVD)\cite{Bonaccorso_mt_12,Bae_nn_10} and transferred on top of the DBRs with SiO$_2$-coating as described in Methods. The quality of graphene before and after transfer is monitored by Raman spectroscopy\cite{Ferrari_prl_06,Cancado_nl_2011,Ferrari_nn_13}. The Raman spectrum before transfer is shown in Fig.\ref{figure2}(a). This is measured at 457nm excitation, since this wavelength suppresses the Cu luminescence, which would result in a non-flat background\cite{Lag_apl_13}. The spectrum shows a very small I(D)/I(G)$\sim$0.004, indicating negligible defects\cite{Ferrari_prl_06,Cancado_nl_2011,Ferrari_nn_13,Ferrari_prb_00}. The 2D peak is a single sharp Lorentzian with full width at half maximum, FWHM(2D)$\sim$35cm$^{-1}$, a signature of SLG \cite{Ferrari_prl_06}. Representative Raman spectra of the transferred graphene on the 0, $\lambda$/4 SiO$_{2}$, $\lambda$/8 SiO$_{2}$, $\lambda$/12 SiO$_{2}$ devices are shown in Fig.\ref{figure2}(a). After transfer, the 2D peak is still a single sharp Lorentzian with FWHM(2D)$\sim$35cm$^{-1}$, confirming that SLG has indeed been successfully transferred, and I(D)/I(G)$\sim$0.005, showing that negligible additional defects are induced by the transfer process. In order to estimate the doping level of the transferred films, an analysis of more than 10 measurements is carried out for 514nm excitation. We use this wavelength as most previous literature and correlations were derived at 514nm\cite{Das_nn_08}. For the film transferred on the $\lambda$/8 SiO$_{2}$ sample, the average G peak position, Pos(G), and FWHM(G), are 1591.8cm$^{-1}$ and 14.6cm$^{-1}$. The average Pos(2D) is 2693cm$^{-1}$, and the 2D to G intensity and area ratios I(2D)/I(G); A(2D)/A(G), are 2.7 and 6.3. This indicates a p-doping$\sim$0.5$\times$10$^{13}$cm$^{-2}$, corresponding to a Fermi level shift$\sim$300meV\cite{Das_nn_08}. Similarly, for the $\lambda$/4- and $\lambda$/12-mirrors we get a p-doping$\sim0.8\times 10^{13}$cm$^{-2}$, corresponding to a Fermi level shift<400meV. For comparison, we also transferred SLG on quartz, Fig.\ref{figure2}(b). The band at$\sim$0.270$\mu$m is a signature of the van Hove singularity in the graphene density of states\cite{kravets_prb_2010}, while those at$\sim$1.4, 2.2$\mu$m are due to the quartz substrate \cite{Lag_apl_13}. The absorption at 960nm (our operation wavelength) is$\sim2.3\%$, but it decreases to$\sim$ 1\% at 2$\mu$m due to doping\cite{Mak_prl_08,Lag_apl_13}. By fitting to the measured transmittance ($T_{\text{r}}\approx 1- A$), we get $E_{\text{F}}\sim350$meV, consistent with the Raman estimates. However, such doping level is not enough to significantly affect the absorption at 960nm, which is measured to be$\sim2.3\%$ (Fig.\ref{figure2}(b)), as for intrinsic SLG\cite{Nair_s_08,Mak_prl_08}.

The linear unsaturated absorption of our four GSAMs at 960nm, measured with a high-precision (0.05\% resolution) reflectivity setup\cite{Maas_oe_08} is plotted in Fig.\ref{figure1}(e). This also plots the calculated absorption from Eq.\ref{equationFE_abs}. Our devices have A=0.25\%, 1.6\%, 3.2\% and 10\% at 960nm respectively, in agreement with calculations. Fig.\ref{figure1}(f) plots the field intensity enhancement calculated from Eq.\ref{equationFE_abs} as a function of wavelength, compared to experiments. This further validates the results. Note that the absorption is not flat as that of graphene on quartz (Fig.\ref{figure2}(b)), because $\xi$ depends on wavelength (Fig.\ref{figure1}(f)) according to Eq.\ref{equationFE_abs}.

We also characterize the GSAMs reflectivity as a function of input light fluence (J/cm$^{2}$). The fluence-dependent reflectivity measurements (non-linear reflectivity) show an increase in reflectivity with fluence as expected from a SA, Fig.\ref{figure2}(d). The maximum changes in reflectivity for the $\lambda$/12, $\lambda$/8 and $\lambda$/4 devices are 0.2\%,0.9\% and 2\%, respectively. The measurement for the $\lambda$/4 SiO$_2$ device (\textit{i.e.} the sample with $\xi$=4 at the graphene position) is shown in Fig.\ref{figure2}(c). For a fast SA (i.e. where the absorber recovery time is faster than the probe pulse duration), the reflectivity can be written as\cite{Lee_apb_2012}:
\begin{equation} \label{equation4}
R(F) \approx \frac{R_{\text{lin}}-R_{\text{ns}}}{\sqrt{\frac{F}{F_{\text{sat}}}+\frac{F}{F_{\text{sat}}}^2}} \text{atanh}\left[\sqrt{\frac{F}{F_{\text{sat}}+F}}\right]+R_{\text{ns}},
\end{equation}
where $R_{\text{lin}}$ is the unsaturated reflectivity, $R_{\text{ns}}$ the non-saturable reflectivity, $F_{\text{sat}}$ the saturation fluence. We estimate a saturation fluence $F_{\text{sat}}\sim$100$\mu$J/cm$^2$ (corresponding to a peak intensity $I_{\text{peak}}\sim$1.0GW/cm$^2$), as extracted by fitting Eq.\ref{equation4} to the data in Fig.\ref{figure2}(c). The estimated modulation depth is$\sim5\%$, 2.7 times larger than that reported for SLG on quartz\cite{Baek_ape_2012}. When a higher input fluence (>120$\mu$J/cm$^2$ (4GW/cm$^2$)) is used, the GSAM reflectivity starts to increase permanently, indicating degradation. From Eq.\ref{equation4}, $F_{\text{sat}}$ of the $\lambda$/8  sample is estimated as$\sim$200$\mu$J/cm$^2$, higher than the $\lambda$/4 sample, because the smaller field intensity enhancement at the absorber makes the device saturate at a higher fluence. In this case, degradation also starts at higher fluence (>300$\mu$J/cm$^2$). In SLG, the non-equilibrium (non-thermal) distribution of electrons in conduction band and holes in valence band created by an ultrafast pulse relaxes, eventually reaching thermal equilibrium with the lattice, due to various processes\cite{Brida_nc_13,Tomadin_prb_13}, including carrier-carrier and carrier-phonon scattering, as well as radiative electron-hole recombination (non-linear photoluminescence\cite{Liu,heinz,Bonaccorso_np_10}). In the sub-ps time-frame two main processes occur: first, the initial peak produced by the pump laser broadens, due to carrier-carrier collisions, converging towards a hot Fermi-Dirac shape on an ultrashort time scale<100fs\cite{Brida_nc_13,Tomadin_prb_13}. On a longer timescale, optical phonon emission\cite{lazzeri} drives a cooling in which the Fermi Dirac distribution shifts towards the Dirac point\cite{malic12,Brida_nc_13,Tomadin_prb_13}.

For VECSEL mode-locking we select the $\lambda$/8 GSAM because it offers suitable linear loss ($<$3\%). This device also provides a larger modulation depth ($>0.9\%$) compared to the $\lambda$/12 GSAM. The laser cavity configuration is sketched in Fig.\ref{figure3}(a), with a picture in Fig.\ref{figure3}(b). The resonator mode and pump spot radius on the gain chip are 150$\mu$m. In order to achieve a sufficient intensity to saturate the GSAM, we implement a beam waist$\sim$30$\mu$m on the absorber using a concave folding mirror with a 20mm radius of curvature. A picture of the $\lambda$/8-GSAM is shown in Fig.\ref{figure3}(c).

We obtain stable mode-locking with a pulse duration of 466fs (Fig.\ref{figure4}(a)). The spectrum is centered at$\sim$949nm with FWHM=2.5nm, Fig.\ref{figure4}(b). Note that the field intensity enhancement of our $\lambda$/8 GSAM is $\xi_{\text{abs}}$=1.5 at 949nm (compared to 1.3 at 960nm, Fig.\ref{figure1}(f)). The pulse repetition rate is 2.5GHz, detected with a fast photodiode and measured with a microwave spectrum analyzer, see Fig.\ref{figure4}(c,d), one order of magnitude higher than previous fiber\cite{Sun_an_10,Sun_pe_12} and solid-state\cite{Lag_apl_13,Sun_pe_12} lasers mode-locked by graphene, due to the compactness of our VECSEL design. The time-bandwidth product is 0.353, 1.1 times larger than what expected for transform-limited sech$^2$ pulses, indicating that the output pulses are slightly chirped (i.e. the instantaneous frequencies are time-dependent\cite{Saleh_book}). The average output power is 12.5mW, with a 0.2\% output coupling transmission. Higher power up to 26mW with 2ps pulses is also achieved using a 0.5\% OC transmission. We calculate the input pulse fluence on the GSAM as$\sim$125$\mu$J/cm$^2$, corresponding to a reflectivity modulation$\sim$0.55\%, according to Fig.\ref{figure2}(a).

In order to verify the broadband operation of our GSAM, we also perform a wavelength-tuning study using the VECSEL described above and an additional quantum-well (QW) VECSEL optimized for emission at$\sim$970nm. We use a$\sim$10cm cavity at 1.5GHz, with various OC transmission rates and gain chips to fully test our GSAMs. A Fabry-P\'{e}rot fused silica etalon (20$\mu$m thick) is used for wavelength tuning. In order to optimize the output power at a given emission wavelength, the gain chip heat sink temperature is adjusted between -20 and +20$^o$C. Mode-locked operation is obtained in a range from 935 to 981nm (46nm), with pulse durations up to 8ps (Fig.\ref{figure5}(d)). Figs.\ref{figure5}(a,b) show the pulse duration and average output power for different emission wavelengths. A maximum tuning range of 21nm with a single VECSEL gain chip is achieved with the 970nm QW VECSEL, Fig.\ref{figure5}(c). This is larger than previously reported with any SESAM mode-locked VECSEL\cite{Morris_pw_12}.

In conclusion, we demonstrated a versatile approach to engineer the absorption of graphene saturable absorber mirrors in the 0-10\% range. Accordingly, the saturation fluence can be adjusted with the field intensity enhancement. We mode-locked VECSELs with a series of different gain chips in a wavelength range as broad as 46nm (from 935 to 981nm) with repetition rate up to 2.48GHz and 466fs pulse duration. This can lead to novel graphene based ultrafast light sources to meet the wavelength range, repetition rate and pulse duration requirements for various applications (e.g. metrology, spectroscopy and data-communication).
\section*{Methods}
\subsection*{Mirror preparation}
30-pair anti-resonant AlAs/GaAs (81.1nm/67.85nm) DBRs are grown on 600$\mu$m thick GaAs by molecular beam epitaxy (MBE, VEECO GEN III). Subsequently, the wafer is cleaved into 1$\times$1 cm$^2$ pieces and different SiO$_2$ coatings are deposited using a plasma enhanced chemical vapor deposition reactor (Oxford Instruments PECVD 80+). The layer thickness is measured on reference Si samples with an ellipsometer (SENTECH SE850).
\subsection*{GSAM preparation}
SLG is grown by CVD\cite{Bae_nn_10} by heating a 35$\mu$m thick Cu foil to 1000$^o$C in a quartz tube, with 10sccm H$_2$ flow at$\sim$5$\times$10$^{-2}$ Torr. The H$_2$ flow is maintained for 30min in order to reduce the oxidized Cu surface\cite{Bonaccorso_mt_12,Bae_nn_10} and to increase the graphene grain size\cite{Bonaccorso_mt_12,Bae_nn_10}. The precursor gas, a H$_2$:CH$_4$ mixture with flow ratio 10:15, is injected at a pressure of 4.5$\times$10$^{-1}$ Torr for 30min. The carbon atoms adsorb onto the Cu surface and form SLG via grain propagation\cite{Bonaccorso_mt_12,Bae_nn_10}. The quality and number of graphene layers are investigated by Raman spectroscopy (Renishaw InVia micro-Raman spectrometer equipped with a Leica DM LM microscope and a 100X objective)\cite{Ferrari_prl_06,Cancado_nl_2011,Ferrari_nn_13}. A 5$\times$5mm$^2$ SLG is transferred onto the mirrors as follows\cite{Lag_apl_13,Bonaccorso_mt_12}: First, a layer of poly(methyl meth-acrylate) (PMMA) is spin-coated on the samples. The Cu foil is etched using a mixture of 3\% H$_2$O$_2$: 35\% HCl (3:1 ratio), which is further diluted in equal volume of deionized water. The PMMA/graphene films are then rinsed in two consecutive deionized H$_2$O baths. Next, the films are picked up on the mirror substrates and left to dry under ambient conditions. Finally, the PMMA is dissolved in acetone, leaving the SLG films on the mirrors. The transferred SLG is inspected by optical microscopy, Raman spectroscopy, and absorption spectroscopy. The non-linear reflectivity of the GSAMs is measured using the high-precision reflectivity setup described in Ref.\cite{Maas_oe_08}. A Kerr-lens mode-locked Ti:Sapphire laser (Tsunami, Spectra-Physics) is used as a probe laser, with 100fs pulse duration at a 80MHz repetition rate, with$\sim$740mW average power at 960nm.
\subsection*{VECSEL laser and characterization}
QW VECSELs emitting at$\sim$940 and 970nm are grown by metal-organic vapor phase epitaxy (MOVPE, AIXTRON AIX 200/4) as for Ref.\cite{Lorenser_apb_04}. A QD VECSEL with an emission wavelength$\sim$950nm is grown by MBE as described in Ref.\cite{Hoffmann_oe_11}. Instead of 9 QD layers placed in 7 subsequent anti-nodes of the electric field as in Ref.\cite{Lorenser_apb_04}, the gain chip we use here has 2 QD layers placed in the first anti-node, whereas no QDs are placed in the 6$^\text{th}$ anti-node to balance the stronger excitation due to higher absorption of the pump light around the first anti-nodes. All gain structures are grown in reverse order, and subsequently processed on a diamond heat sink grown by CVD, purchased from Diamond Materials GmbH, as described in Ref.\cite{Haring_jqe_02}. The pump laser is coupled into a 200$\mu$m fiber. The laser output is characterized using a RF spectrum analyzer (HP8592L, Agilent 8565EC) and a fast photodiode (New Focus 1434). An optical spectrum analyzer (HP 70952) is used to detect the optical spectrum. The pulse train is temporally characterized with an intensity autocorrelator (Femtochrome FR103XL).
\subsection*{Field intensity enhancement}
The absorption of a graphene layer on the top of a mirror is defined by the field intensity enhancement at the absorber position. This is determined by the constructive and destructive interference of the electric field of the incident and reflected beam. To calculate the field intensity enhancement, we assume an incident optical wave:
\begin{equation}
\mathcal{E}_{\text{in}}\left({z}\right)=\mathcal{E}_{\text{in}}^0 e^{i\left({\omega t-k_nz}\right)},
\end{equation}
where $k_n=2\pi n/\lambda$ is the wave number in the material, n is the refractive index of the material in which the light is propagating. The reflected optical wave is $\mathcal{E}_{\text{out}}\left({z}\right)=-\mathcal{E}_{\text{in}}\left({-z}\right)$, because of total reflection at the mirror, the node at the surface of the mirror and the propagation in the opposite direction. From Eq.\ref{equationFEdef}, we get the field enhancement $\xi$ for an anti-resonant high-reflection mirror with no additional coating in air:
\begin{equation}
\begin{split}
\xi\left({z}\right)&=\frac{\left|\mathcal{E}_{\text{in}}\left({z}\right) -\mathcal{E}_{\text{in}}\left({-z}\right) \right|^2}{\left|\mathcal{E}_{\text{in}}\left({z}\right)\right|^2}=\left|2i\sin\left(k_nz\right)\right|^2\\
&=4\sin\left(\frac{2\pi n_{\text{air}} z}{\lambda}\right).
\end{split}
\end{equation}
Then we consider the field enhancement of an anti-resonant high-reflection mirror with a SiO$_2$-coating of thickness $d$. At the air-SiO$_2$-interface we have\cite{Saleh_book}:
\begin{equation}\label{eq_r}
r_{\text{in}}=\frac{n_{\text{air}}-n_{\text{SiO}_2}}{n_{\text{air}}+n_{\text{SiO}_2}}=\frac{1-n_{\text{SiO}_2}}{1+n_{\text{SiO}_2}} \text{ and }\\r_{\text{out}}=-r_{\text{in}},
\end{equation}
where $r_{\text{in}}$ and $r_{\text{out}}$ are the Fresnel coefficients\cite{Saleh_book} of reflection at normal incidence at the air-SiO$_2$ and SiO$_2$-air interface. The corresponding Fresnel coefficients\cite{Saleh_book} for transmission at the air-SiO$_2$ and SiO$_2$-air interface are:
\begin{equation}\label{eq_t}
\begin{split}
t_{\text{in}}&=\frac{2n_{\text{air}}}{n_{\text{air}}+n_{\text{SiO}_2}}=\frac{2}{1+n_{\text{SiO}_2}}\\
t_{\text{out}}&=\frac{2n_{\text{SiO}_2}}{n_{\text{air}}+n_{\text{SiO}_2}}=\frac{2n_{\text{SiO}_2}}{1+n_{\text{SiO}_2}}.
\end{split}
\end{equation}
The electric field of the reflected beam consists of the superposition of the incoming beam ($\mathcal{E}_{\text{air}}^{\text{in}}$) reflected at the air-SiO$_2$ interface, and the electric field of the beam ($\mathcal{E}_{\text{SiO}_2}^{\text{out}}$) transmitted in SiO$_2$ at the same interface:
\begin{equation}\label{eqI}
\mathcal{E}_{\text{air}}^{\text{out}}=r_{\text{in}}\mathcal{E}_{\text{air}}^{\text{in}}+t_{\text{out}}\mathcal{E}_{\text{SiO}_2}^{\text{out}},
\end{equation}
whereas the electric field of the incident beam in SiO$_2$ at the interface is:
\begin{equation}\label{eqII}
\mathcal{E}_{\text{SiO}_2}^{\text{in}}=t_{\text{in}}\mathcal{E}_{\text{air}}^{\text{in}}+r_{\text{out}}\mathcal{E}_{\text{SiO}_2}^{\text{out}}
\end{equation} and the electric field of the reflected beam in SiO$_2$ at the interface is:
\begin{equation}\label{eqIII}
\mathcal{E}_{\text{SiO}_2}^{\text{out}}=r_{\text{mirror}}e^{2i n_{\text{SiO}_2}k_0d}\mathcal{E}_{\text{SiO}_2}^{\text{in}}.
\end{equation}
From Eqs.\ref{eqII},\ref{eqIII} we get with normalization of the incoming field ($\mathcal{E}_{\text{air}}^{\text{in}}=1$):
\begin{equation}
\mathcal{E}_{\text{SiO}_2}^{\text{in}}=\frac{t_{\text{in}}}{1+r_{\text{out}} e^{2i n_{\text{SiO}_2}k_0d}}
\end{equation} and
\begin{equation}\label{eqIV}
\mathcal{E}_{\text{SiO}_2}^{\text{out}}=\frac{-t_{\text{in}}e^{2i n_{\text{SiO}_2}k_0d}}{1+r_{\text{out}} e^{2i n_{\text{SiO}_2}k_0d}}.
\end{equation}
Inserting Eq.\ref{eqIV} in Eq.\ref{eqI} we get the electric field of the reflected beam in air:
\begin{equation}\label{eqV}
\mathcal{E}_{\text{air}}^{\text{out}}=r_{\text{in}}+t_{\text{out}}\frac{-t_{\text{in}}e^{2i n_{\text{SiO}_2}k_0d}}{1+r_{\text{out}} e^{2i n_{\text{SiO}_2}k_0d}}.
\end{equation}
inserting Eqs.\ref{eqV},\ref{eq_r},\ref{eq_t} in Eq.\ref{equationFEdef} we get:
\begin{widetext}
\begin{equation}
\xi(d_{\text{SiO}_2})=\left|1-\frac{4n_{\text{SiO}_2}}{(1+n_{\text{SiO}_2})^2}\frac{1}{e^{2i n_{\text{SiO}_2}k_0d}+\frac{n_{\text{SiO}_2}-1}{n_{\text{SiO}_2}+1}}+\frac{1-n_{\text{SiO}_2}}{1+n_{\text{SiO}_2}}    \right|^2
\end{equation}
\end{widetext}
which gives:
\begin{equation}
\xi(d_{\text{SiO}_2})=\frac{4}{1+n_{\text{SiO}_2}^2\cot^2\left(k_0 n_{\text{SiO}_2} d_{\text{SiO}_{2}}\right)}
\end{equation}

\section*{Acknowledgments}
We thank Prof. T. S\"{u}dmeyer for useful discussions. We acknowledge funding from the Royal Society, the European Research Council Grant NANOPOTS, EU grants RODIN, GENIUS, MEM4WIN, CareRAMM, and Graphene Flagship (contract no. NECT-ICT-604391), EPSRC grants EP/K01711X/1,\linebreak EP/K017144/1, EP/G042357/1, Nokia Research Centre, Emmanuel College, Cambridge, the FIRST clean room facility of ETH and the Swiss National Science Foundation (SNSF).

\end{document}